\documentclass{desyproc}

\begin{document}
\title{\vspace{-3cm}{\small \hfill{DESY 11-223; MPP-2011-139}}\\[1.8cm]
Solar Hidden Photon Search}

\author{{\slshape Matthias Schwarz$^1$, Axel Lindner$^2$, Javier Redondo$^3$, Andreas Ringwald$^2$, G\"unter Wiedemann$^1$}\\[1ex]
$^1$Hamburger Sternwarte, Gojenbergsweg 112, D-21029 Hamburg, Germany\\
$^2$Deutsches Elektronen-Synchrotron DESY, Notketra\ss e 85, D-22607 Hamburg, Germany\\ 
$^3$Max-Planck-Institut f\"ur Physik, F\"ohringer Ring 6, D-80805 M\"unchen, Germany}

\contribID{schwarz\_matthias}

\desyproc{DESY-PROC-2011-04}
\acronym{Patras 2011} 
\doi  

\maketitle

\begin{abstract}
The Solar Hidden Photon Search (SHIPS) is a joint astroparticle project of the Hamburger Sternwarte and DESY. The main target is to detect the solar emission of a new species of particles, so called Hidden Photons (HPs). Due to kinetic mixing, photons and HPs 
can convert into each other as they propagate. 
A small number of solar HPs -- originating from photon $\to$ HP oscillations in the interior of the Sun -- can be converted into 
photons in a long vacuum pipe pointing to the Sun -- the SHIPS helioscope.
\end{abstract}

\section{Introduction}

Hidden Photons (HPs) are the gauge bosons of a hypothetical hidden local U(1) symmetry. Such symmetries arise in popular extensions of the Standard Model, especially in those based on string theory~\cite{arXiv:1002.1840}. Known particles have no direct interaction with HPs (hence the latter are hidden), but still HPs may have a tiny residual interaction with them, as very massive particles with both electric and hidden 
charge can generate kinetic mixing with the standard photon~\cite{Holdom:1985ag}. In this case, the natural value of the kinetic mixing angle $\chi$ is that of a quantum correction, $\chi\sim e e_h/(16\pi^2)$. Since the hidden gauge coupling $e_h$ can be very small and because of possible cancellations between different mediator contributions, there is no clear minimum for $\chi$: values in the $10^{-16}$---$10^{-3}$ range have been predicted in the literature~\cite{Dienes:1996zr,Abel:2008ai,Goodsell:2009xc,Cicoli:2011yh,Goodsell:2011wn}. The very feeble interaction makes HPs promising candidates for the dark sector that current cosmology and astrophysics are revealing. They have been proposed as Dark Matter (DM)~\cite{Redondo:2008ec,Pospelov:2008jk,Nelson:2011sf,Arias:Patras} 
and as mediating Dark Forces between DM particles~\cite{ArkaniHamed:2008qn,arXiv:1110.2636}. 
Moreover, if their mass is in the meV and their kinetic mixing in the micro range, 
their cosmological relic abundance could also provide the right amount of extra 
dark radiation~\cite{Jaeckel:2008fi} favored by recent CMB observations~\cite{Komatsu:2010fb,arXiv:1009.0866,Keisler:2011aw}.

The kinetic mixing term induces flavor oscillations between photons and HPs~\cite{Okun:1982xi}. 
After a propagation length $L$, a HP has a probability to convert into a photon (or viceversa) given by
\begin{equation}
P(\gamma'\leftrightarrow \gamma )= \frac{\sin^2 2\chi}{\left(\cos 2\chi +f\right)^2+\sin^2 2\chi}
\sin^2\left(\frac{m_{\gamma'}^2 L \sqrt{\left(\cos 2\chi +f\right)^2+\sin^2 2\chi}}{4\omega}\right),
\label{osc_prob}
\end{equation}
where $m_{\gamma'}$ is the HP mass, setting the oscillation length $\propto 4\omega/m_{\gamma^{\prime}}^2$, with $\omega$ the photon frequency, 
and $f=2\omega^2\Delta n/m_{\gamma'}^2$ encodes medium effects via $\Delta n= n-1$, with $n$ the photon index of refraction in the medium. 

\begin{wrapfigure}{r}{0.45\textwidth}
\centerline{\includegraphics[width=0.45\textwidth]{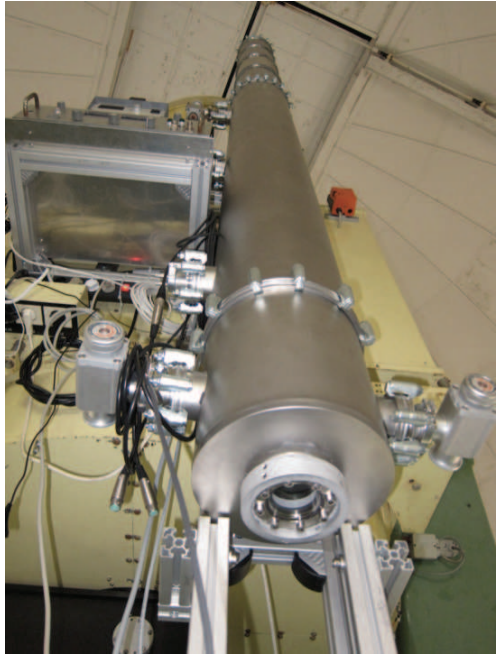}}
\caption{The helioscope TSHIPS attached to the equatorial mount of the Oskar L\"uhning Telescope at the Hamburger Sternwarte.}
\label{Fig:TSHIPS}
\end{wrapfigure}

This oscillation mechanism enables us to search for HPs even if $\chi$ is tiny. 
One of the most powerful means to search for HPs is to measure the flux of HPs from the Sun~\cite{Redondo:2008aa,Gninenko:2008pz}. 
These HPs originate from oscillations of photons from the outer layers of the Sun's interior and their flux predictions 
are covered by one of us in another article in these proceedings~\cite{HPflux}. 
The enhanced luminosity from the solar surface produces a ring-like structure in the signal on Earth that can be used to optimize the signal-to-noise~\cite{Cadamuro:2010ai}. Therefore, imaging detectors might be very advantageous for searching for HPs.

In this article we report on the status of the Solar Hidden Photon Search (SHIPS), a search for the reconversion of solar HPs into detectable photons in a long light-tight vacuum tube tracking the Sun.   
SHIPS is an offspring of the ongoing ALPS (Any Light Particle Search) project~\cite{arXiv:0905.4159,arXiv:1004.1313} at DESY in Hamburg.

\section{Experimental setup}

SHIPS is a helioscope-type experiment very much like SUMICO~\cite{hep-ex/9805026} at Tokio University or the 
CERN Axion Solar Telescope (CAST)~\cite{arXiv:0810.4482}. There is however an important difference: unlike these  
other helioscopes, SHIPS does not exploit a magnet. In fact, a background electromagnetic field is not needed for solar HP search, 
since HP-photon oscillations happen already in vacuum. This is in contrast to axion $\leftrightarrow$ photon oscillations, 
which occur only in electromagnetic fields.    

The expected number of photons originating from the reconversion of solar HPs in the vacuum tube is proportional to the HP flux 
($\Phi_{\gamma^\prime}$) the collecting area ($A$), the tube's length ($L$), 
measuring time ($T$), and the oscillation probability~(\ref{osc_prob}),  
\begin{equation}
\label{Ngamma}
N_\gamma = A T \int \frac{d \Phi_{\gamma'}}{d\omega} P(\gamma'\to \gamma)d\omega .
\end{equation}
The oscillation probability can be enormously suppressed if $f\gg 1$, i.e. if the index of refraction is large. Therefore, the gas density has to be kept under a certain minimum given by the HP mass.  
To obtain optimal results, the collecting area and tube length should be as large as possible.

The current Telescope for Solar HIdden Photon Search TSHIPS consists of a 430 cm long stainless steel tube with a diameter of 25 cm combined from the tubes of two prototype vacuum vessels and a detector compartment (see Fig.~\ref{Fig:TSHIPS}). The upper tube is a lightweight vault structure developed for this project. TSHIPS is mounted on the Oskar L\"uhning Telescope (OLT) located at the Hamburger Sternwarte in Hamburg-Bergedorf 
and can thus be remotely controlled.
It will be run in a pressure range of $10^{-6}$~mbar or lower. A membrane pump creates a pre-vacuum of $10^{-2}$~mbar. A turbopump, directly attached to TSHIPS, establishes the required minimal pressure of less than $10^{-4}$~mbar in the helioscope volume of more than 260 liters within minutes.

Regenerated electromagnetic photons propagate along the same trajectories as their HP progenitors. This enables us to observe a source like our sun in the `light' of HPs. 
The optical system is placed inside the detector compartment. The principal item is a Fresnel lens with a focal length of about 20 cm. This device ensures a high transitivity and image quality in the optical and near infrared band,   
where the solar flux of sub-eV mass HPs dominates as they are created deep in the solar photosphere. 
Initially available ultra-low-noise detectors are two
photomultiplier tubes (PMTs) and a low-noise CCD camera, all cooled to minimize dark current. 
The frequency range of the detectors is chosen to be optical and near infrared. 
TSHIPS is currently mounted with a PMT featuring a dark current of approximately 2.8 Hz at ambient temperature. 
This makes single photons detectable.
The other available PMT has even a lower dark current of 0.5 Hz.

\begin{figure}[h]
\centerline{\includegraphics[width=0.6\textwidth]{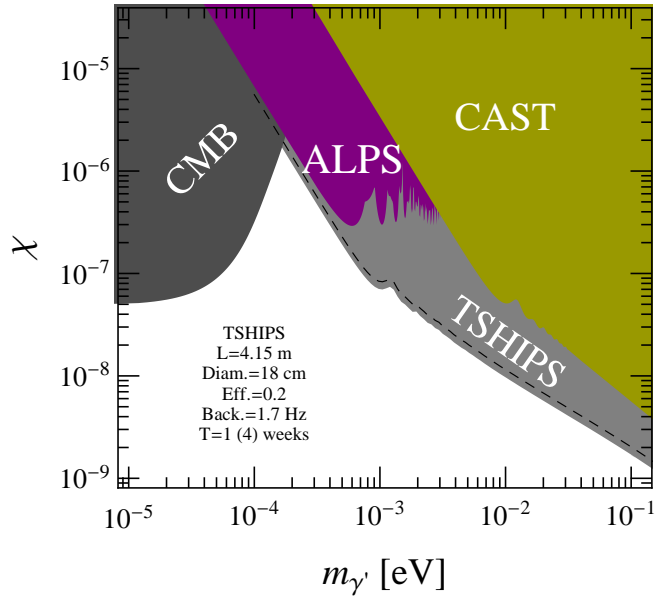}}
\caption{Current exclusion regions for hidden photons with kinetic mixing angle $\chi$ and mass $m_{\gamma'}$, in particular from 
the light-shining-through-a-wall experiment ALPS and from the helioscope experiment CAST, 
together with the projected sensitivity of TSHIPS (gray region).}
\label{Fig:parspace}
\end{figure}

\section{Present status and outlook}

The first measurements and results on HPs are expected within the next weeks. 
We estimated the dark current noise to be $\gamma_{\rm dc}\simeq 1.7$ Hz. 
The significance of an HP flux discovery scales as $S=2\left(\sqrt{N_\gamma+\gamma_{\rm dc}T}-\sqrt{\gamma_{\rm dc}T}\right)\sim N_\gamma/\sqrt{\gamma_{\rm bc}T}$.
Figure~\ref{Fig:parspace} shows the expected $S<3$ exclusion limit (similar to 95\% C.L.) achievable in 1 and 4 weeks of Sun tracking. Assuming a hypothetical flux of solar HPs producing one photon every 100 seconds on TSHIPS, a discovery would be achievable during a data taking period of less than 2 days. 
The physical values of HP mass and coupling parameter could then be estimated by tuning experimental parameters in Eq.~(\ref{Ngamma}) such as the refractive index or the oscillation length.  

Phase II planning has started for a much wider (diameter of 125 cm) and longer (13~m) helioscope. 
This massive tube will be put on a separate large alt az mount for long term operation and will probably be installed on the DESY premises. Detector development is ongoing in conjunction with other astronomical programs.

 

\begin{footnotesize}

\end{footnotesize}


\end{document}